\documentclass[]{spie}  

 \usepackage{newtxtext,newtxmath}
\usepackage[T1]{fontenc}
\usepackage{ae,aecompl}
\usepackage{multirow}
\usepackage{multicol}
\usepackage{rotating}
\usepackage{lscape} 
\usepackage{longtable}
\usepackage{graphicx}	
\usepackage{xcolor}
\usepackage{rotating}
\usepackage{tabularx}
\usepackage{lscape} 
\usepackage{arydshln}
\usepackage{comment}
\usepackage{amsmath,amsfonts}
\usepackage[colorlinks=true, allcolors=blue]{hyperref}
\graphicspath{{./}{figures/}}

\title{Twinkle - a small satellite spectroscopy mission for the next phase of exoplanet science}

\author[1]{Ian Stotesbury}
\author[1]{Billy Edwards}
\author[2]{Jean-Francois Lavigne}
\author[3]{Vasco Pesquita}
\author[2]{James J. Veilleux}
\author[1]{Philip Windred}
\author[1]{Ahmed Al-Refaie}
\author[1]{Lawrence Bradley}
\author[1]{Sushuang Ma}
\author[1]{Giorgio Savini}
\author[1]{Giovanna Tinetti}

\author[5,6]{Til Birnstiel}
\author[7,8]{Sally Dodson-Robinson}
\author[5,6]{Barbara Ercolano}
\author[4]{Dax Feliz}
\author[9]{Scott Gaudi}
\author[4]{Nina Hernitschek}
\author[10]{Daniel Holdsworth}
\author[11]{Ing-Guey Jiang}
\author[12]{Matt Griffin}
\author[13]{Nataliea Lowson}
\author[5,6]{Karan Molaverdikhani}
\author[14,15]{Hilding Neilson}
\author[9]{Caprice Phillips}
\author[5]{Thomas Preibisch}
\author[12]{Subhajit Sarkar}
\author[4]{Keivan G.\ Stassun}
\author[10]{Derek Ward-Thompson}
\author[13]{Duncan Wright}
\author[16]{Ming Yang}
\author[17]{Li-Chin Yeh}
\author[16]{Ji-Lin Zhou}

\author[1]{Richard Archer}
\author[1]{Yoga Barrathwaj Raman Mohan}
\author[1]{Max Joshua}
\author[1]{Marcell Tessenyi}
\author[1]{Jonathan Tennyson}
\author[1]{Benjamin Wilcock}

\affil[1]{Blue Skies Space Ltd., 69 Wilson Street, London, UK}
\affil[2]{ABB, 800 Bd Hymus, Saint-Laurent, Canada}
\affil[3]{Airbus Defense and Space, Gunnels Wood Rd, Stevenage, UK}
\affil[4]{Department of Physics and Astronomy, Vanderbilt University, Nashville, TN 37235, USA}
\affil[5]{Universit\"ats-Sternwarte, Fakult\"at f\"ur Physik, Ludwig-Maximilians-Universit\"at M\"unchen, Scheinerstr.~1, 81679 M\"unchen, Germany}
\affil[6]{Exzellenzcluster `Origins', Boltzmannstr.~2, 85748 Garching, Germany}
\affil[7]{Department of Physics and Astronomy, University of Delaware, 217 Sharp Lab, Newark, DE 19716, USA}
\affil[8]{Bartol Research Institute, Sharp Lab, 104 The Green, Newark, DE, 19716, USA}
\affil[9]{Department of Astronomy, The Ohio State University, 100 W 18th Ave, Columbus, OH 43210, USA}
\affil[10]{Jeremiah Horrocks Institute, University of Central Lancashire, Preston PR1 2HE, UK}
\affil[11]{Department of Physics and Institute of Astronomy, National Tsing Hua University, Hsin-Chu, Taiwan}
\affil[12]{School of Physics and Astronomy, Cardiff University, Cardiff, CF24 3AA, UK}
\affil[13]{Centre for Astrophysics, University of Southern Queensland, Centre for Astrophysics, 499-565 West Street, Toowoomba, QLD 4350, Australia}
\affil[14]{Dept of Physics \& Physical Oceanography, Memorial University of Newfoundland \& Labrador}
\affil[15]{David A. Dunlap Dept of Astronomy \& Astrophysics, University of Toronto}
\affil[16]{School of Astronomy and Space Science, Nanjing University, Nanjing 210023, Jiangsu, China}
\affil[17]{Institute of Computational and Modeling Science, National Tsing Hua University, Hsin-Chu, Taiwan}

\authorinfo{Send correspondence to Ian Stotesbury (E-mail: ian@bssl.space)}

\begin{document} 
\maketitle

\begin{abstract}

With a focus on off-the-shelf components, Twinkle is the first in a series of cost competitive small satellites managed and financed by Blue Skies Space Ltd. The satellite is based on a high-heritage Airbus platform that will carry a 0.45 m telescope and a spectrometer which will provide simultaneous wavelength coverage from 0.5–4.5 $\rm{\mu m}$. The spacecraft prime is Airbus Stevenage while the telescope is being developed by Airbus Toulouse and the spectrometer by ABB Canada. Scheduled to begin scientific operations in 2025, Twinkle will sit in a thermally-stable, sun-synchronous, low-Earth orbit. The mission has a designed operation lifetime of at least seven years and, during the first three years of operation, will conduct two large-scale survey programmes: one focused on Solar System objects and the other dedicated to extrasolar targets. Here we present an overview of the architecture of the mission, refinements in the design approach, and some of the key science themes of the extrasolar survey.
 
\end{abstract}



\section{Introduction}
\label{sec:intro}  

Twinkle, the first mission from Blue Skies Space Ltd. (BSSL), will deliver high quality science data outside of the typical space agency model. Through directly financing construction, BSSL aims to bolster the provision of science data by building on advances in the space sector thanks to “New Space” satellite delivery models.  Our innovative model leverages heritage space industry products to re-purpose them for scientific application, turning science mission concepts into reality on accelerated timescales \cite{archer_bssl}. In doing so, BSSL will deliver cost-effective scientific satellites that can be complementary to both flagship ground and space-based observatories, an approach that is in line with recommendations by scientists \cite{boley_new_space, serjeant_small_sat}.

Twinkle, due to being operations in 2025, is the first of these spacecraft and makes use of this approach to utilise heritage components where possible to deliver the spacecraft. Twinkle has a designed operational lifetime of at least seven years and time on the spacecraft will be available to scientists through multi-year global survey programmes or via dedicated time for the user. Mission operations will commence with two surveys running in parallel over three years: an extrasolar survey and a solar system survey. These surveys will deliver visible and infrared spectroscopy of thousands of objects within and beyond our solar system, providing homogeneous datasets that will enable in-depth meticulous studies of populations of objects. The survey programme will allow both experienced and early career participating scientists to collaborate and produce world-leading research in planetary, exoplanetary and stellar science.

Here we present the Twinkle mission, summarising the work done by the Airbus and ABB industrial teams working with Blue Skies Space Ltd (BSSL) and highlighting the technical and programmatic updates that have occurred since the publication of previous papers outlining the mission \cite{savini,jason,edwards_twinkle_exo,edwards_twinkle_ast,edwards_twinkle_solar}. Additionally, we summarise some of the key science interests of the current members of Twinkle's extrasolar survey.

\section{Design Summary}

Recent work as part of an industrialisation phase have meant the performance specifications of the spacecraft have been refined and updated since previous publications detailing the design \cite{savini,jason}. The chief design refinements are in the pointing methodology employed, the resolving power of the instrument, and the consolidation of the optical design down to a single array. A summary of the key design parameters, and any changes from the previous baseline, are detailed in Table \ref{design_overview}. We note that these design parameters have been utilised for some time to model the performance of the mission \cite{edwards_terminus}, but that this is the first formal manuscript to have detailed these developments. In the following sections we provide more detailed information on some major elements of the mission's design and operations.

\begin{table}[h]
\centering
\resizebox{0.95\textwidth}{!}{
    \begin{tabular}{c|ccc|cc}
    \multirow{2}{*}{Parameter}& \multicolumn{3}{c|}{Old Design} & \multicolumn{2}{c}{New Design} \\ 
     & Channel 0 & Channel 1 & Channel 2 & Channel 0 & Channel 1 \\ \hline
    Wavelength range & 0.45 $\rm{\mu m}$ - 1.0 $\rm{\mu m}$ & 1.3 $\rm{\mu m}$ - 2.43 $\rm{\mu m}$ & 2.43 $\rm{\mu m}$ - 4.5 $\rm{\mu m}$ & 0.5 $\rm{\mu m}$ - 2.43 $\rm{\mu m}$ & 2.43 $\rm{\mu m}$ - 4.5 $\rm{\mu m}$ \\
    Resolving Power & 250 & 250 & 60 & up to 70 & up to 50 \\
    Detector Temperature & 250 $K$ & 70 $K$ & 70 $K$ & \multicolumn{2}{c}{90 $K$} \\
    Detector manufacturer & \multicolumn{3}{c|}{Selex} & \multicolumn{2}{c}{Teledyne} \\
    Pointing method & \multicolumn{3}{c|}{FGS} & \multicolumn{2}{c}{High-Performance Gyroscope}\\
\end{tabular}}
\caption{Summary of key design parameters and changes from feasibility study \cite{savini,jason}.}
\label{design_overview}
\end{table}

\subsection{Spacecraft System Overview}

The Twinkle design has been developed around the heritage spacecraft bus `AstroBus S', a repackaging of the `AS250' which has a design lifetime of 10 years. The spacecraft bus is suited for rapid construction and has previously been used for PeruSat\footnote{www.airbus.com/space/earth-observation/perusat.html}, while also being adapted for the CHEOPS mission \cite{benz_cheops}. The spacecraft is designed for a 700 km, sun-synchronous dawn-dusk orbit, to maintain an orbit normal aligned closely to the anti-sun direction for the telescope. The Field of Regard (FoR), the cone of potential attitudes the spacecraft can adopt for observations, is centred on the anti-Sun vector with a maximum displacement of $\pm$40$^\circ$. As well as the Sun exclusion constraint, the FoR is constrained throughout the seasons by Earth obstruction in part of the FoR for part of the orbit. The modelling of this, and the subsequent impact on science, is discussed in Section \ref{orbital_tool}.

The AstroBus platform requires adaptation for Twinkle in the mechanical design to provide a Sun shield of sufficient size. The spacecraft makes use of sunshields to maintain the cool thermal environment by shading the telescope body and payload bay from both Sun and Earth flux across the wide FoR. The cryogenic temperatures are reached by a combination of space-facing radiators, which are thermally attached to the telescope interface panel, and active coolers with a dedicated radiator thermally attached to the inner sanctum and detector.

The spacecraft's sole science instrument is a spectrometer operating from 0.5 to 4.5 $\rm{\mu m}$, using two spectral channels (with the split at 2.43 $\rm{\mu m}$) which are re-imaged to a single Teledyne H2RG series detector. The telescope and its optical elements will be maintained at less than 160 K and the primary aperture of the telescope will be greater than 0.45 m. The design is being developed by Airbus and ABB, making use of heritage components where possible as well as the product lines and supply chains of both companies. The inner sanctum of the spectrometer is to be cooled to lower than 100 K and the detector maintained below 90 K with a target temperature of below 85 K.

Twinkle's Low Earth Orbit (LEO) enables the use of an X-band downlink capable of downloading all payload data via an isotropic antenna. This will allow the spacecraft to maintain scientific operations during downlink sessions and not constrain spacecraft attitude. A wide variety of ground stations offer downlink access for satellites operating in a Sun-synchronous orbit and these options continue to be explored by BSSL.

\begin{figure}
    \centering
    \includegraphics[width=\columnwidth]{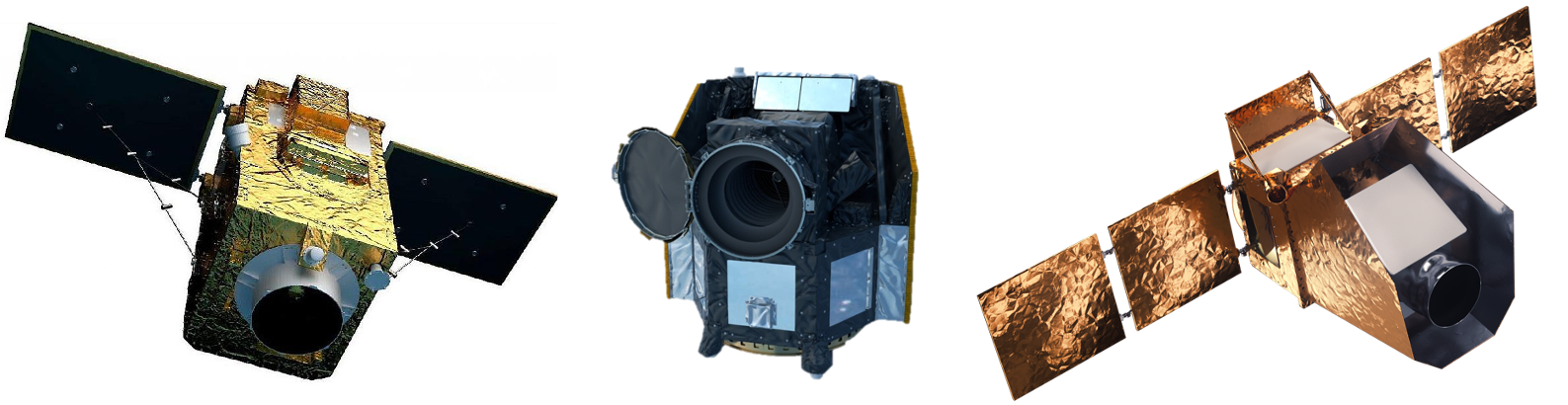}
    \caption{PeruSat (left, image credit: Airbus Defence and Space), CHEOPS (middle, image credit: ESA) and Twinkle (right, image credit: BSSL) all utilise the Astrobus S platform. While the primary mirror of CHEOPS has a diameter of 0.3 m, Twinkle will carry a 0.45 m primary offering over twice the collecting area. The Twinkle telescope assembly will come from Airbus's NAOMI product range, which was also used for Perusat.}
    \label{fig:my_label}
\end{figure}

\subsection{Thermal Solution}
\label{thermal}

There are three key temperatures that drive the scientific performance of the mission: the temperatures of the telescope and optical chain, the inner instrument sanctum and the detector. The primary sources of flux to reject are the Sun and the Earth and a number of configurations were considered for the mission with an important consideration being the FoR allowance of each thermal solution. 

A Sun-synchronous orbit was selected to help constrain the Earth-Sun-Spacecraft geometry and hence the impact of Earth flux on the spacecraft thermal design. A number of Sun shields designs were considered with the telescope either side-mounted or top-mounted relative to the spacecraft bus. No significant thermal preference between these two designs approaches was identified so, for mechanical considerations, the top-mounted orientation was selected.

The implementation of an all passive thermal solution would offer the lowest cost, and complexity, approach for a mission such as Twinkle. While the analysis showed the passive solution to be potentially viable, there was insufficient margin to take forward as the baseline and instead the decision was made to make use of an active cooler to control the temperature of the detector. The industrial team have performed a thorough analysis of the available low cost, low lift, cryocoolers, shortlisting several as well as determining a preferred option. The cryocooler has been incorporated into the design in order to ensure a low operating temperature for the detector as well as providing the temperature stability need to achieve the required for scientific performance.

\subsection{Optical Design}

The spacecraft optical design has been developed with the ABB team to minimise the complexity and to consider industrialisation. Key developments are the change from using narrow science and background slits to a wider field stop instead. The optical paths of the channels have been consolidated onto a single detector, as opposed to two separate spectrometers, with now only two diffracting elements as opposed to three. An additional benefit of this approach is the simplification of the thermal isolation considerations for the optical bench.

Each optical path consists of 6 mirrors, one dichroic and a grism element to produce the spectra. The detector is baselined as a Teledyne H2RG series array, which will be actively cooled to a achieve a maximum temperature of 90 K (with a design target of < 85K). The telescope assembly will come from Airbus's New AstroSat Optical Modular Instrument (NAOMI) product range and the primary mirror will have a diameter of at least 0.45 m. The mirrors will be cooled below 168 K while the inner sanctum will be cooled to have maximum of operating temperature of 110K (with target temperatures of <158 K and <100 K, respectively).

The new optical design reduces the complexity of the mission by allowing for one detector and a single optical plane. In the previous design, the incorporation of the UVIS detector, while a heritage solution, introduced a required operating temperature that was much higher than the rest of the optical bench \cite{savini,jason}. The electrical interface and performance of Teledyne detectors are well known, having been widely tested and often used for astrophysics applications by Airbus and ABB as well as the wider industry. 

\begin{figure}
    \centering
    \includegraphics[width=0.45\columnwidth]{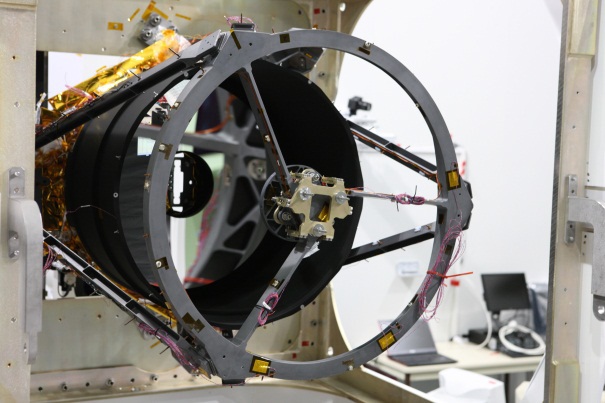}
    \includegraphics[width=0.46\columnwidth]{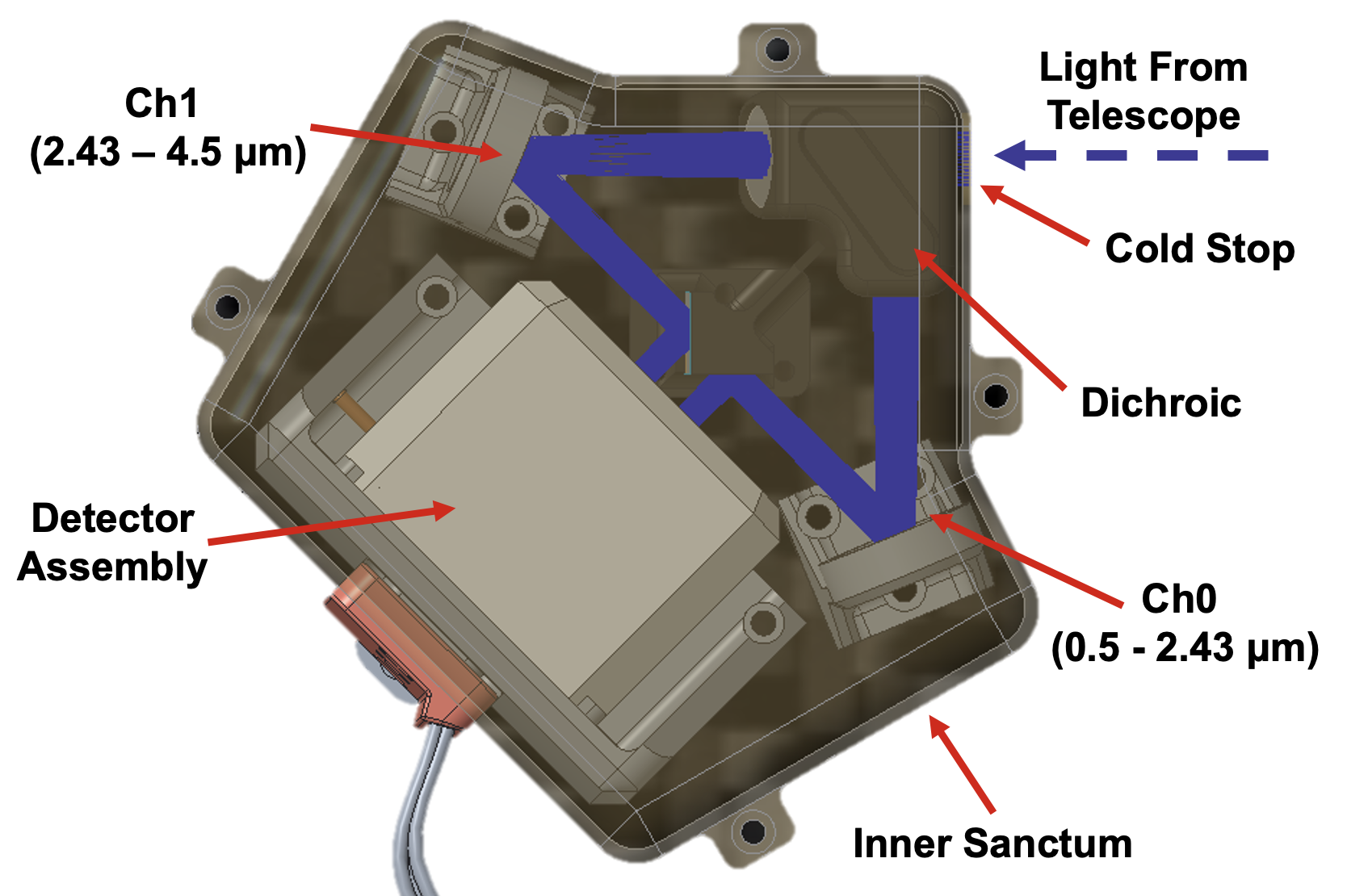}
    \caption{Left: NAOMI telescope product as used for the PeruSat mission (image credit: Airbus Defence and Space). Right: Spectrometer concept (image credit: ABB). The spectrometer is located in the inner sanctum and it interfaces optically with the telescope at the cold stop. Two separate channels, with different dispersions, are re-imaged on the same detector to provide the simultaneous 0.5-4.5 $\mu$m spectral coverage.}
    \label{fig:spectrometer}
\end{figure}

The spectrometer is located in the inner sanctum which operates at a lower temperature than the telescope. It interfaces optically with the telescope at its output pupil where a cold stop will be used to minimize thermal background contamination. A dichroic filter will reflect the spectral region between 0.5 and 2.43 $\rm{\mu m}$ and will transmit the light between 2.43 and 4.5 $\rm{\mu m}$. Each of the two channels has its own grism to adapt the dispersion to their specific science needs. Fold mirrors are then used to image the two spectral channels onto the same detector. The basic layout of the spectrometer is shown in Figure \ref{fig:spectrometer}. A field stop in this design is located prior to the cold stop, at an intermediate focal plane within the telescope, to limit the contamination from field stars.  The integrated payload design of the telescope and spectrometer is now being refined at system level to select the field stop diameter, cold stop diameter and the plate scale on the detector to achieve the best photometric stability.  

\subsection{Wavelength Coverage}

As highlighted above, the updated spectrometer design now maintains two spectral channels, with a continuous wavelength coverage from 0.5 to 4.5 $\rm{\mu m}$ and a split at 2.43 $\rm{\mu m}$. The continuous coverage across the wavelength range is a distinct improvement on the previous design where some light (1.0 - 1.3 $\rm{\mu m}$) was sequestered to the FGS. The previous design also maintained three separate channels covering 0.4 - 1.0 $\rm{\mu m}$, 1.3 - 2.43 $\rm{\mu m}$, and 2.43 - 4.5 $\rm{\mu m}$, with two separate detectors required. 

A combination of considerations led the short wavelength cutoff to be relaxed, including detector and mirror performance and heritage, to maintain a cost-effective mission. Multiple telescope providers and manufacturing processes were considered and each presented similar issues. The wavelength range was relaxed to remove this driver to the spacecraft design after the change was considered against the scientific impact and shown to be acceptable. Similarly, the 4.5 $\mu$m cut-off was determined as a comprise between the additional science that can be achieved by covering longer wavelengths and improving the SNR across the whole wavelength range. The long wavelength cut-off for the chosen detector can go as far as 5.3 $\mu$m. However, adopting a shorter wavelength cut-off reduces both the dark current and the sensitivity to thermal emission from the telescope. Furthermore, having conducted a review of available components, the wavelength coverage was also shown to align well with the performance of suitable grisms. 

\subsection{Pointing Performance}
\label{jitter}

The pointing performance of the spacecraft is a key parameter that determines the viable resolving power of the instrument as well as jitter noise. The industrial team reviewed the science requirements of the mission against the previous pointing performance and the spacecraft buses available. Using Power Spectral Density functions (PSDs) provided by Airbus for the modified AstroBus-S designs, Cardiff University and BSSL conducted simulations for a number of exoplanetary targets using TwinkleSim (see Section \ref{inst_per}). The time-domain aspect of this tool allows for the consideration of pointing jitter noise and the effects of spreading the spectral image on the detector. In order to account for margins, a conservative PSD was taken against which to model and review science performance. 

The analysis was run across a range of detector read-out apertures and for various scaled versions of the PSDs to explore the sensitivity of the mission performance to pointing jitter noise. The investigation highlighted and confirmed the relationship between resolving power and high-frequency jitter. The analysis showed that a high-performance gyroscope will be sufficient to suppress high-frequency pointing noise in the system.

\section{Performance Modelling}
\label{tools}

BSSL aims to create satellites which are driven by scientific performance. To enable the design of Twinkle to be optimised, BSSL's software team have developed a suite of tools to model the performance of the spacecraft and its instrumentation. These are being actively used to design the spacecraft, as well as being utilised by researchers to explore how Twinkle can contribute to their scientific interests.

\subsection{Instrument Performance}
\label{inst_per}

Two different types of software have been created to model the performance of Twinkle: a radiometric code and a time-domain simulator. The radiometric model, TwinkleRad, is based upon the open-source code ExoRad, a generic radiometric tool that has also been adapted for the ESA Ariel mission \cite{arielrad,tinetti_ariel,tinetti_ariel2}. It accounts for a variety of noise sources including shot noise, instrument emission, dark current, read noise and zodiacal light. Noise due to pointing fluctuations (i.e. jitter) can be imported from dynamical modelling.

The time-domain simulator, TwinkleSim, is adapted from ExoSim \cite{Exosim} which models noise on a frame-to-frame basis, with the targeted specification for the Twinkle instrument. Each exposure in TwinkleSim consists of up-to-ramp non-destructive reads. The output covers the source photon noise, dark current noise, zodiacal emission noise, telescope and instrument emission noise, read noise, spatial jitter noise and spectral jitter noise. TwinkleSim is now being updated to NextGenSim, which is using the new generation of code architectures to optimise the numerical performance and code structure to support the intensive computational techniques need to conduct high-frequency, long-duration, time-dependent optical chain and detector simulations. NexoGenSim will provide a comprehensive simulation of Twinkle's science performance, starting from the light source and yielding realistic scientific data products.

\subsection{Target Availability}
\label{orbital_tool}

Observatories in low Earth orbits can experience interruptions in target visibility due to Earth occultations. Additionally, instruments and spacecraft usually have specific target-Sun, target-Moon or target-Earth limb restrictions. To account for these, BSSL has developed an orbital tool \cite{edwards_terminus} which is capable of modelling the orbit of a spacecraft. Additionally, the tool can calculate the angle between the target and the Earth limb, the Sun, or any other celestial body, in a similar way to tools used for other missions (e.g. for CHEOPS \cite{kuntzer_salsa}). 

Twinkle will operate in a polar, Sun-synchronous orbit, with an altitude of approximately 700 km at an inclination of 90.4 $^\circ$. We previously modelled this geometrically, with the code verified against the mission design analysis and operation software Freeflyer\footnote{\url{https://ai-solutions.com/freeflyer/}}, but the tool has recently been upgraded to utilise two-line element sets (TLEs) using the SGP4 python package\footnote{\url{https://github.com/aholinch/sgp4}}. Hence, Twinkle's orbit is currently modelled using the TLE of CHEOPS \cite{benz_cheops}, a mission which currently operates in a polar, Sun-synchronous orbit at 700 km. The exact altitude of Twinkle's orbit will depend upon the launcher utilised, and the performance of the vehicle during the launch, but this is the target orbit and representative for planning purposes at this stage. The code can impose a number of exclusion angles, with obvious examples being the Sun, Earth and Moon. The first of these is the result of thermal constraints while the latter two are primarily to reduce stray light. The Earth and Moon exclusion angles for Twinkle are under review and conservative values, similar to those of other observatories designed for operation in sun-synchronous orbits \cite{kuntzer_salsa, swain_finesse}, have been used with the agreement of Airbus.

\begin{figure*}
    \centering
    \includegraphics[width=\textwidth]{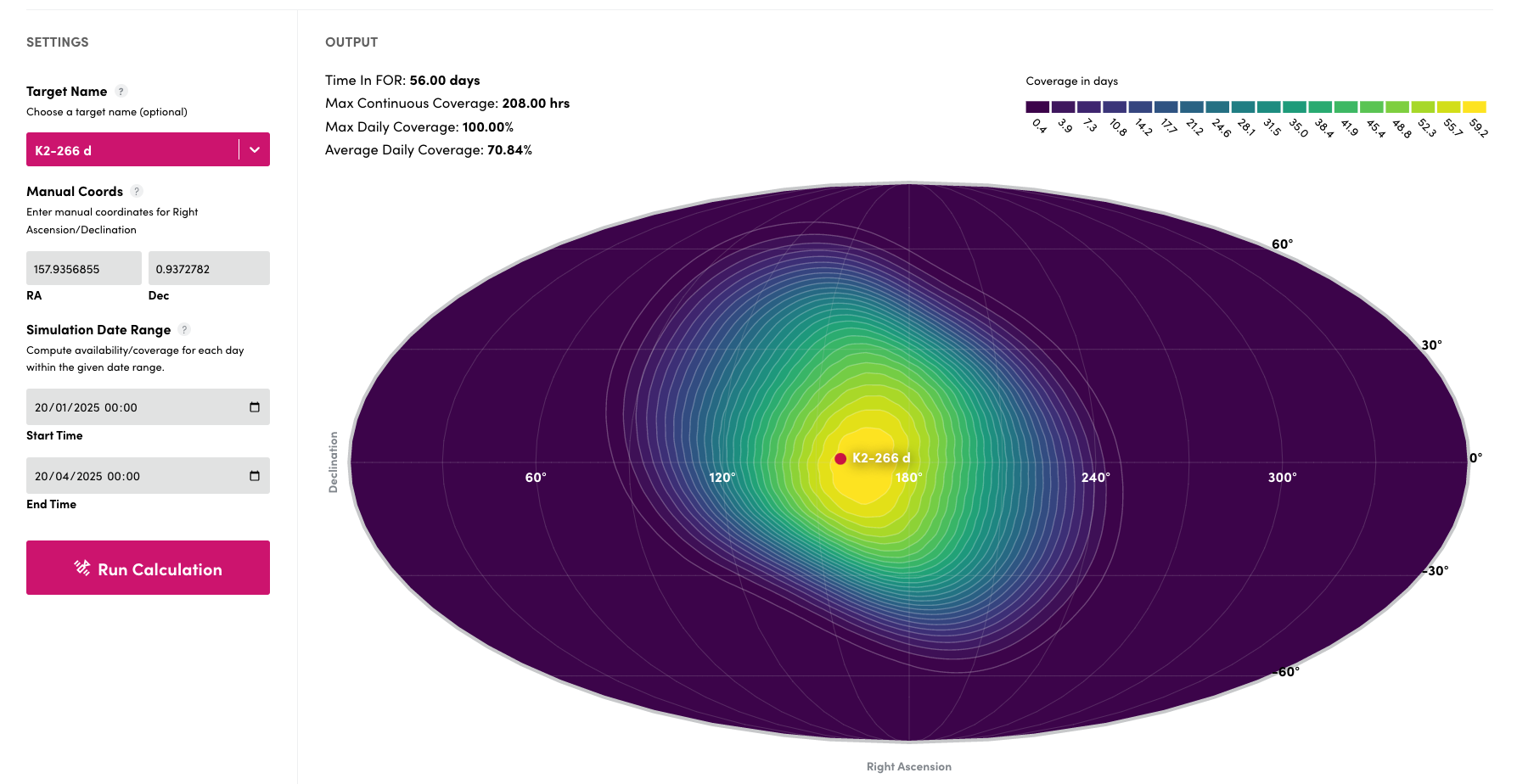}
    \caption{Example output from the orbital tool on Stardrive. K2-266 hosts a multi-planet system and could be a key target for studies of transit timing variations with Twinkle. The accessibility of this target changes across the year, with periods with little to no Earth obscuration existing. }
    \label{fig:sky_vis}
\end{figure*}

\begin{figure*}
    \centering
    \includegraphics[width=\textwidth]{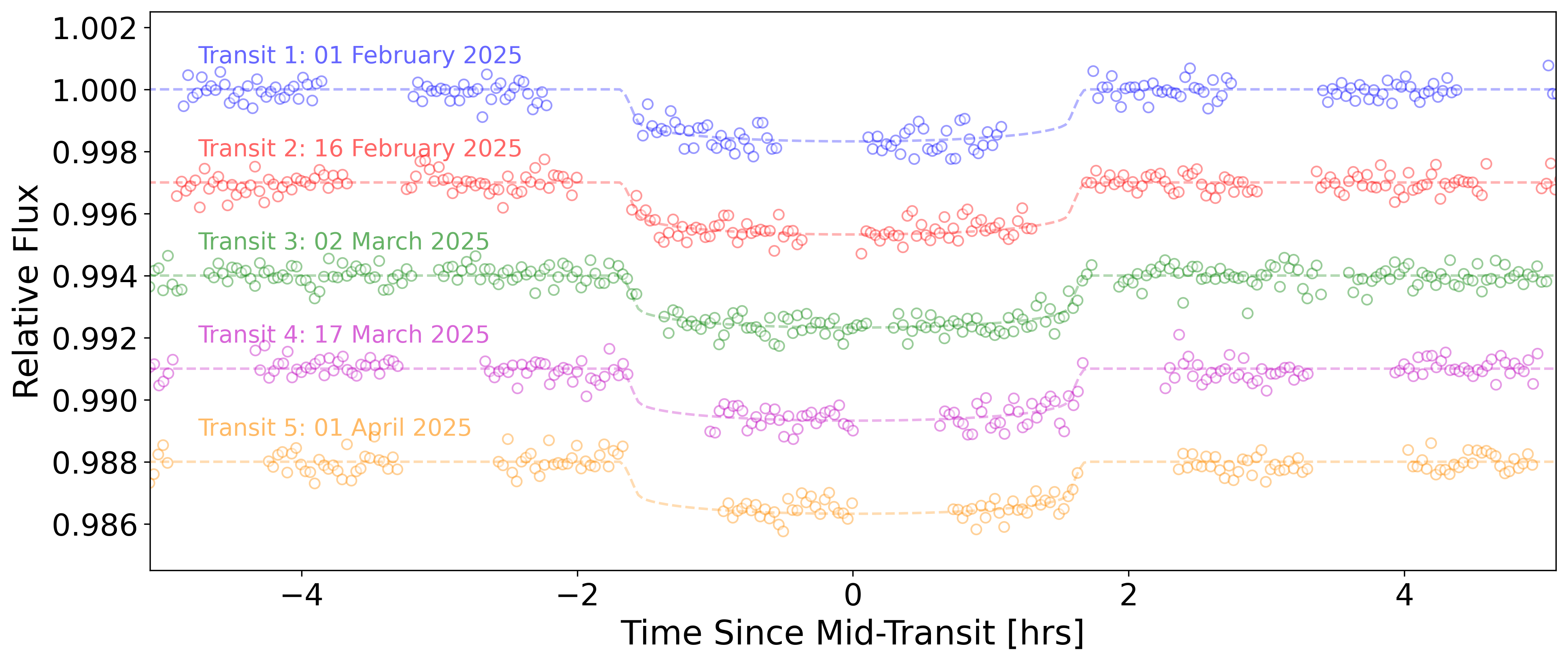}
    \caption{Predicted transits computed within the scheduling tool to allow scientists to identify observations of choice. The target shown in K2-266\,d, a potential target for studies related to transit timing variations, which has five transits which are observable with Twinkle in 2025. Gaps in the light curves are due to Earth obscuration and stray light constraints. For this science case, capturing the ingress and egress of the transits is critical and the tool allows users to understand which observing window(s) are best. TwinkleRad has been used to model the expected white light curve performance for Ch0. The transit light curve was modelled using PyLightcurve \cite{tsiaras_plc} and random Gaussian scatter added to the data points based on the predicted photometric uncertainties. We note that Ch1 will also have a white light curve, captured simultaneously, but this is not plotted here.}
    \label{fig:k2_266_transits}
\end{figure*}

A web-based version of the orbital tool has been created to allow the scientists who are part of Twinkle's surveys to assess the availability of key targets for their science. The tool is available via the Stardrive portal\footnote{\url{https://stardrive.twinkle-spacemission.co.uk}}, a platform built by BSSL to enable members of the surveys to collaboratively work on scientific endeavours related to the Twinkle mission. The tool can allow for a general observability to be calculated, an example of which is demonstrated in Figure \ref{fig:sky_vis} for K2-266, a star which hosts at least five planets \cite{JosephK2-266}. Additionally, for observations of time critical events such as exoplanet transits, specific access times can be queried and the availability provided for that period. An demonstration of this is given for the planet K2-266 d \cite{rodriguez_k2_266}, which has 5 available transits over the year 2025 (see Figure \ref{fig:k2_266_transits}). In this figure, the radiometric tool described in Section \ref{inst_per} has been used to model the white light curve of Ch0 (0.5 - 2.43 $\mu$m).

\subsection{Scheduling}

To accommodate multiple users and science needs, BSSL are developing a scheduling tool to manage the various bookings from users and consolidate the various requests into one schedule that can then be provided to the mission planning tool (MPT) for conversion to telecommands to the spacecraft. In building the schedule, the tool needs to make use of the orbital tool to check observational constraints at high granularity and calculate the total number of transits/observation time required for the target science quality. The request to the radiometric tool will review scientific performance across the orbit, and consider changes in the observing environment (e.g. stray light).

There shall be two ways of using the scheduling tool to book time on Twinkle; time critical and flexible bookings. Time critical bookings are intended to be used by scientists to request explicit observations that must be completed at a specific time slot, such as the close approach of the Apophis asteroid in 2029, or based on periodic events (e.g. an exoplanet transit or eclipse). The scheduling tool will attempt to secure time explicit booking requests first, to maximise the requests that can be scheduled. Other bookings are expected to be more flexible, where the target is selected but the specific access is not prescribed (e.g. the characterisation of a protoplanetary disk \cite{ercolano_twinkle_pah}). In these cases, the user will be asked to enter the desired total duration of observation, total predicted SNR in a specific band or other parameters and the scheduling tool will use that as the input for allocating time. These targets will of course still have some scheduling constraints, such as those due to Sun exclusion angle. These requests will generally be scheduled between time-critical observations and, in doing so, will allow for algorithmic time allocation to maximise science utility, science return, and keep the spacecraft flexible to re-schedule the mission as needed.

In order to manage the various bookings from multiple users and science cases within the science surveys the scheduler will consider relevant priority. Priority will be an important aspect to the efficiency and viability of the scheduling tool and is not as simple as time criticality but will attempt to represent fairness. Users should not expect to be able to utilise only time critical bookings to exploit the system. Flexible bookings explained previously will be given a completion date that constrains their flexibility and user bookings will be reviewed against their portion booked as time critical to understand the impact on fairness. Furthermore, it is expected that, given the rate of discovery of new targets, that scientists would welcome the ability to re-prioritise their time allocation and also remove targets that are identified to be less scientifically interesting. Therefore, an additional degree of flexibility will be needed as well as the ability to quickly rerun the long-term and short-term schedule.

\clearpage

\section{Science with Twinkle}

Twinkle's ability to provide simultaneous spectroscopic coverage between 0.5 and 4.5 $\rm{\mu m}$ is driven by the requirement to characterise exoplanet atmospheres, but also facilitates many other science investigations including study of Solar System objects and protoplanetary discs. During the first three years of operation, two large survey programmes will be implemented. The first of these will focus on extrasolar targets while the second will be dedicated to studying objects within our own Solar System. 

The science programmes of these two surveys are defined entirely by the researchers who join the mission. Each survey will have thousands of hours of telescope time, allowing for structured population-level studies to be undertaken. By allowing scientists to dedicate large amounts of time to individual science cases, Twinkle will enable researchers to answer fundamental questions about the formation and evolution of star and planetary systems in our galaxy. In the run-up to launch, they will collaboratively explore uses for the spacecraft with support provided by BSSL in the form of the Stardrive platform and the tools discussed in Section \ref{tools}.  Here we present some of the science endeavours that Twinkle could contribute to, with a focus on the extrasolar survey and some of the interests of the current members.

\subsection{Extrasolar Survey}

Twinkle will have the capability to acquire data which enables a wide variety of extrasolar science. It will provide high-quality infrared spectroscopic characterisation of the atmospheres of hundreds of bright exoplanets, covering a wide range of planetary types. It will also be capable of providing phase curves for hot, short-period planets around bright stars targets and of providing ultra-precise photometric light curves to accurately constrain orbital parameters, including ephemerides and TTVs/TDVs present in multi-planet systems. 

\textbf{Characterising Proto-planetary Discs:} Twinkle's wide wavelength coverage will allow it to study planetary discs via the infrared excess seen compared to the star's expected spectral energy distribution. These discs are the first stage of planetary formation, so detecting and characterising them provides a crucial snapshot which helps develop our understanding of planet formation processes. Twinkle observations will allow for the characterisation of the composition of these discs, enhancing our understanding of the repartition of elements, the metallicity, the carbon-to-oxygen (C/O) ratio, and constraining the presence of Polycyclic Aromatic Hydrocarbons (PAHs) \cite{seok_pah,ercolano_twinkle_pah}, all of which can act as tracers for formation pathways \cite{oberg_2011,mordasini_2016}.

\textbf{Characterising Stars and Brown Dwarfs:} Data from Twinkle offers the potential to exquisitely characterise stars and brown dwarfs. The Twinkle stellar characterisation working group is exploring how observations from the mission can be used to more precisely determine the fundamental properties of exoplanet host stars, which is ultimately essential to understanding the physical properties of the planets and their atmospheres. Long-term observations covering both the visible and the near-infrared offer the chance to measure the variability of stars, helping to constraint flaring rates and to reveal the presence of star spots and faculae. While Twinkle could collect these data for any stars of interest, these observations would be particularly useful if taken for planet hosting stars as they would help determine the effect the spots and faculae have on transmission spectra of the planet to avoid biasing atmospheric constraints \cite{apai_tlse}. Twinkle could also be utilised to study brown dwarfs, measuring their thermal emission which has been shown to be highly variable as the rotation of these objects brings more and less cloudy regions into view  \cite{apai_bd,biller_bd,bowler_2020}.

\textbf{Transit Timing:} The white light curves from Twinkle's two spectroscopic channels will facilitate precise transit timing measurements. One application of this is the study of transit timing variations (TTVs) due to interactions of bodies in a multi-planet system. Studying these TTVs can reveal the planets' masses and constrain the eccentricities of their orbits, helping us to better understand the planets' internal compositions and the dynamical process of their formation and evolution. One key target of interest is likely to be the K2-266 system, in which TTVs have already been spotted \cite{rodriguez_k2_266}. The Twinkle TTV working group was recently awarded time on CHEOPS (PN: AO3-08, PI: Ing-Guey Jiang) to observe transits of K2-266 d and e. These data will already give better constraints on the interactions between the planets in this system but, when combined with Twinkle data, could yield even more precise results due to the increased baseline. Work is ongoing to identify further targets of interest for this science case. Furthermore, Twinkle data could be utilised to search for orbital decay or precession \cite{jiang_w43,turner_w12,wong_w12}, for which a long baseline is again critical. With Spitzer no longer operating, CHEOPS only providing visible data, and HST or JWST time unlikely to be granted purely for this science case, Twinkle's infrared spectrometer will be unique in its ability to provide measurements of the timing of secondary eclipses which will be useful in distinguishing between decay and precession models \cite{turner_w12, wong_w12}.

\textbf{Atmospheric Characterisation:} By undertaking a structured spectroscopic survey of a multitude of planets, Twinkle will provide the first detailed population studies of exoplanetary atmospheres. These homogeneous datasets will enable the expected chemical trends with bulk parameters to be tested, helping to unravel the complex process of planetary formation and evolution. These studies will build upon the work that has already been done with the Hubble Space Telescope and Spitzer \cite{tsiaras_pop, wallack_spitzer_pop, baxter_spitzer_pop, changeat_emission_pop, keating_hierarchical} as well as those that will be conducted with JWST in the coming years. Through transit and eclipse spectroscopy, Twinkle will be capable of identifying and constraining molecules within the atmospheres of exoplanets \cite{edwards_twinkle_exo}, further enhancing our understanding of these worlds. Twinkle's wavelength coverage will give it access to numerous absorption and emission features. Molecules to which Twinkle is sensitive include H$_2$O and NH$_3$, carbon-bearing species such as CO, CO$_2$, CH$_4$ and HCN, as well as metal oxides and hydrides (e.g. FeH, TiO, VO). 

\begin{figure}
    \centering
    \includegraphics[width=0.9\textwidth]{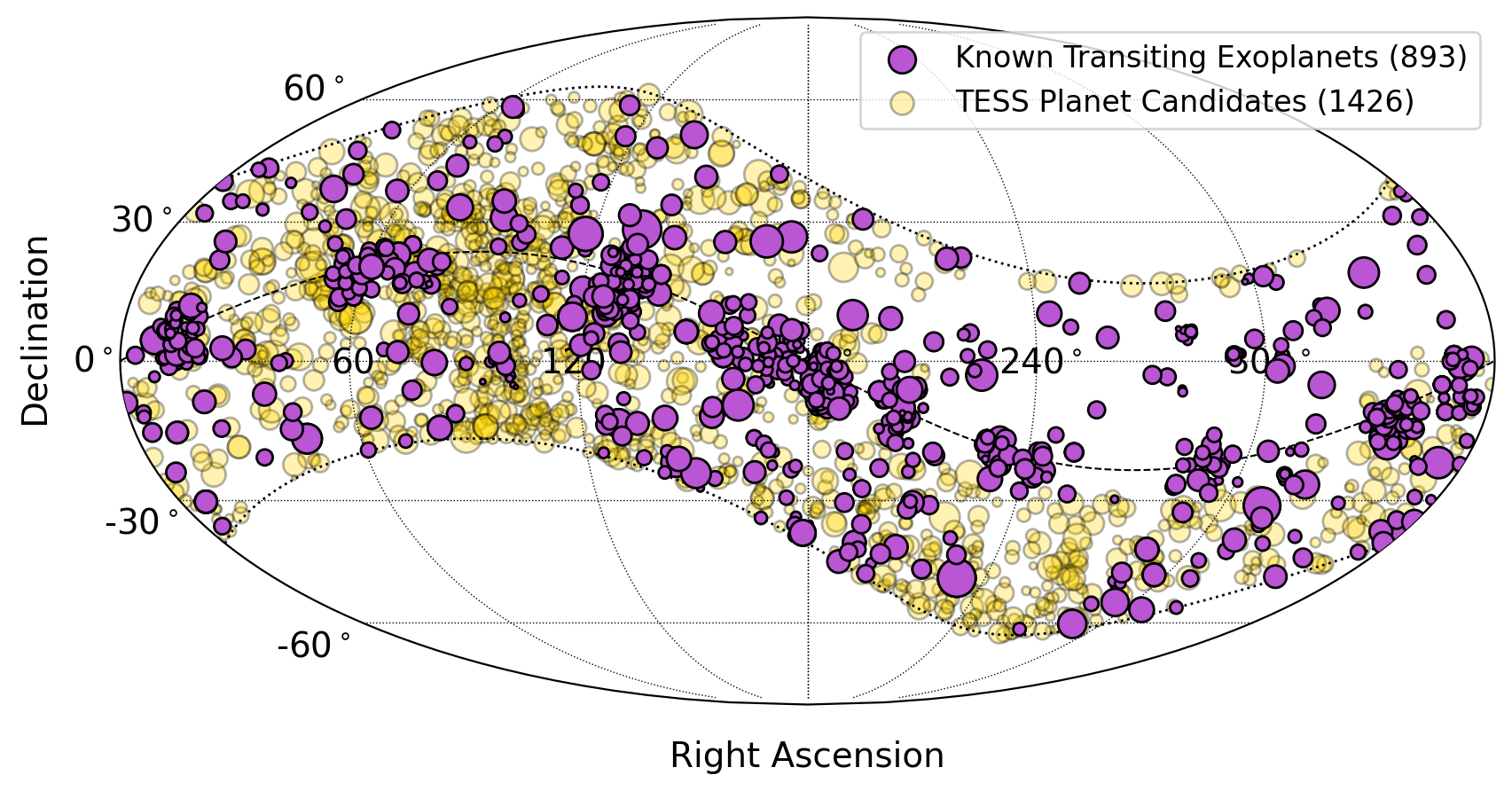}
    \caption{Twinkle can view targets that are within 40 $^\circ$ of the ecliptic plane. Shown here are the currently-known transiting exoplanets. (purple), and planet candidates from TESS (yellow), that lie within Twinkle's Field of Regard (FoR). The size of the points is correlated with the host star's magnitude in the K band, with brighter stars appearing as large circles.}
    \label{fig:sky_loc}
\end{figure}

\begin{figure}
    \centering
    \includegraphics[width=0.477\textwidth]{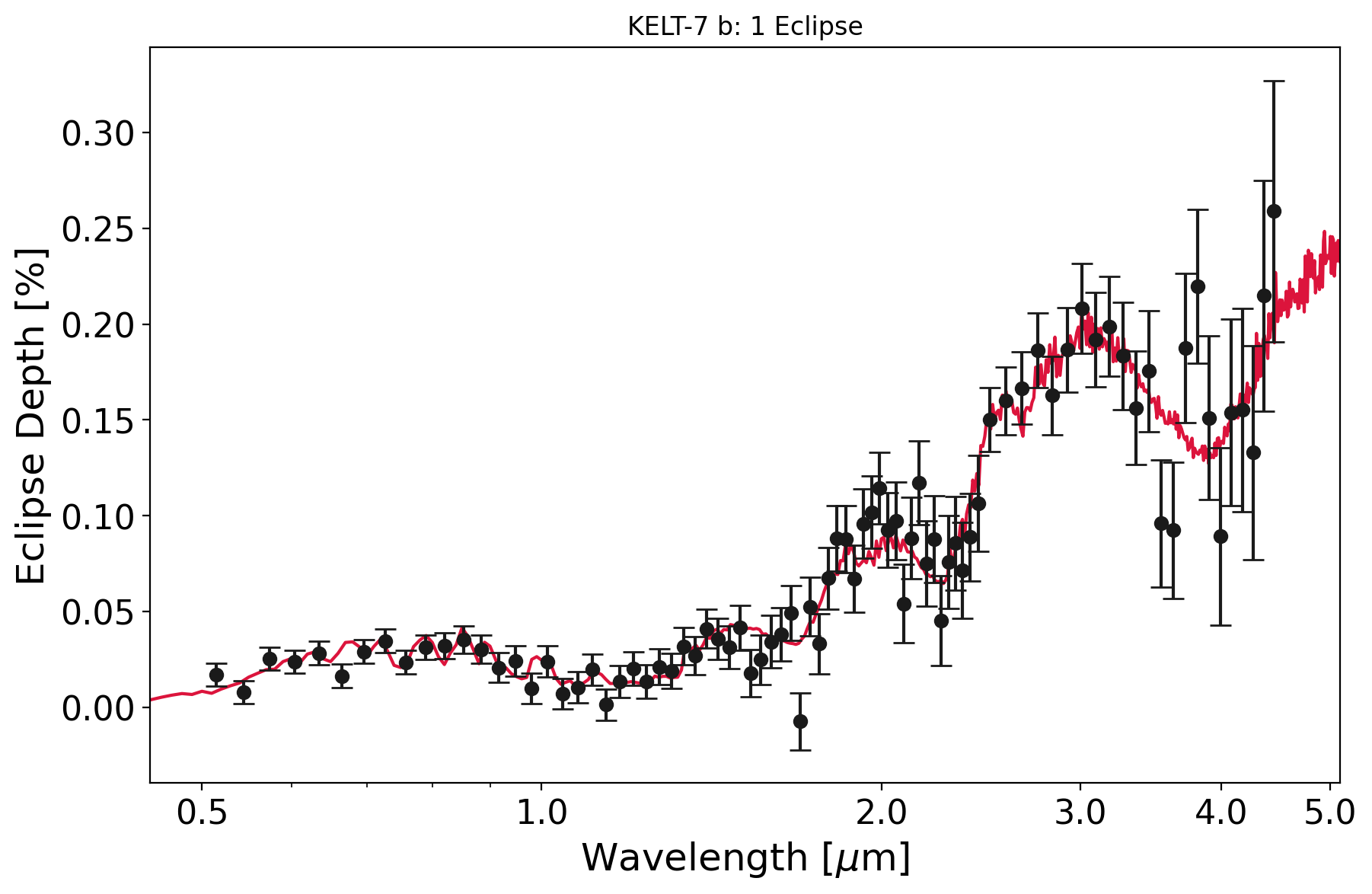}
    \includegraphics[width=0.475\textwidth]{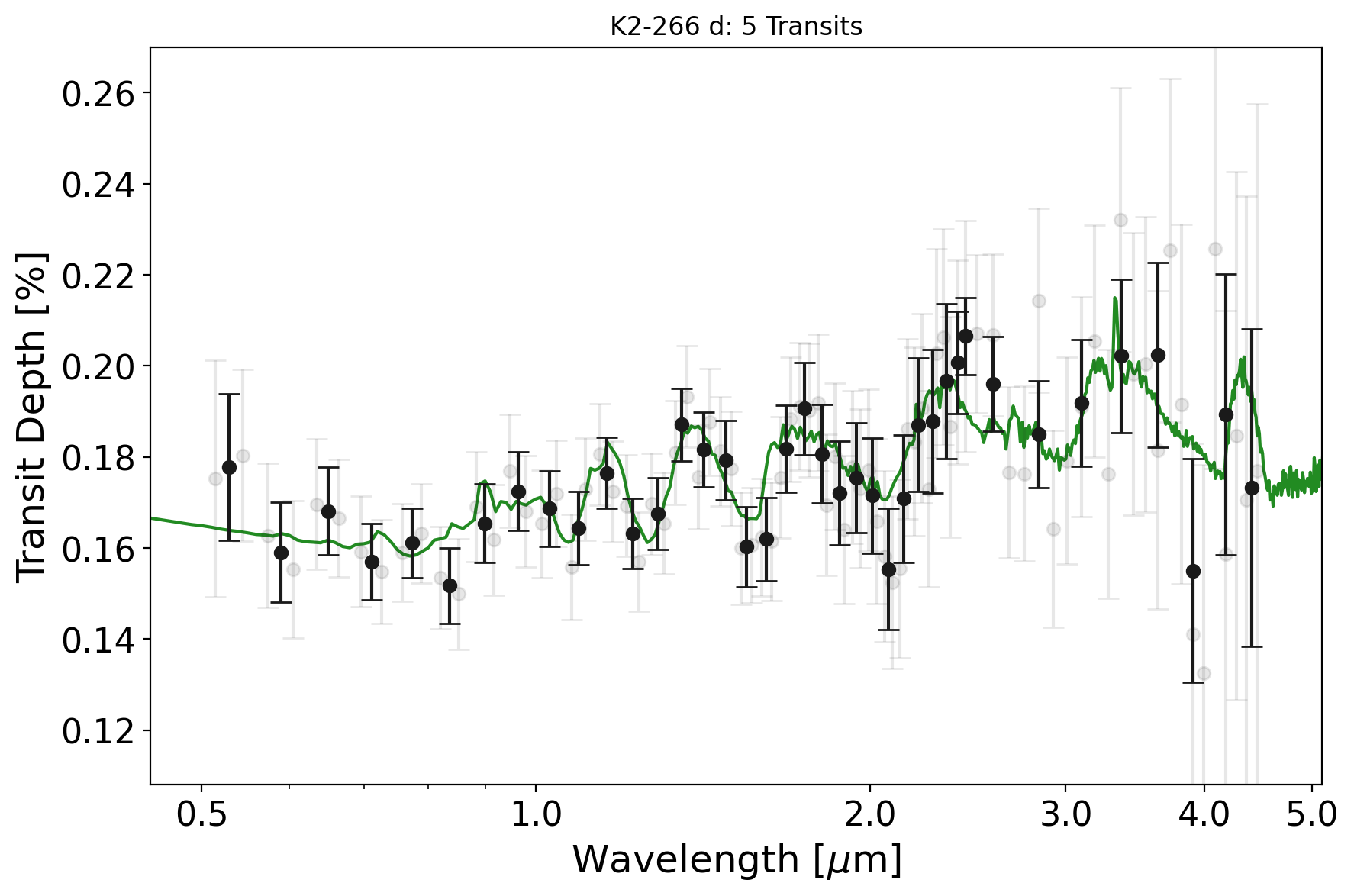}
    \caption{Simulated atmospheric spectra from Twinkle. \textbf{Left:} emission spectrum of KELT-7\,b, based on constraints from current observations\cite{pluriel_k7}, simulated for a single eclipse observation with Twinkle and shown at the standard read-out resolution. Observations of a population of similar planets will allow for studies into potential chemical trends to be undertaken. \textbf{Right:} while K2-266\,d is far from a prime atmospheric target for Twinkle, if transits are taken to study TTVs in the system they will also allow for the atmosphere to be characterised. For this spectrum, five transits are simulated and the data have been spectral binned; the resolution has been halved in Ch0 (0.5\,-\,2.43\,$\mu$m) while three spectral points have been combined for Ch1 (2.43\,-\,4.5\,$\mu$m). The native spectrum, to which Gaussian scatter was added, is shown in grey. Both forward models were produced using TauRex 3 \cite{taurex3} using linelists from \cite{rothman_hitremp_2010,polyansky_h2o,ExoMol_CH4_new,li_co_2015}}.
    \label{fig:spec}
\end{figure}

As shown in Figure \ref{fig:sky_loc}, there are a multitude of exoplanets within Twinkle's FoR and over hundred of these currently-known planets are suitable for study with Twinkle. The many planet candidates from TESS offer further opportunities for atmospheric studies, but Twinkle could also be utilised to validate these by using the white light curves to conduct multi-band photometry. Twinkle will characterise the atmospheres of a diverse population of planets, from temperate Super-Earths to Ultra-Hot Jupiters, with the aim of uncovering chemical trends. As an example, a simulated emission spectrum of the ultra-hot Jupiter KELT-7\,b\cite{bieryla_k7} is given in Figure \ref{fig:spec}. However, each science theme of the Twinkle extrasolar survey is not necessarily distinct and exploiting overlaps between them will help maximise the science yield of the mission. For instance, while K2-266\,d is not a stand-out target for atmospheric characterisation with Twinkle, if transits are taken to study TTVs in the system they may also allow for the atmosphere to be characterised as shown in Figure \ref{fig:spec}.

\textbf{Complementarity with Other Facilities:} Twinkle will be highly complementary to flagship facilities developed by major space agencies. For instance, there will be a significant overlap in the operational timelines for JWST and Twinkle. Given that observing time on JWST will be extremely competitive, Twinkle could be utilised to help identify particularly interesting targets which can then be re-observed with JWST, optimising the usage of its time. Furthermore, some targets will not need the extreme sensitivity that JWST offers and so Twinkle can be utilised to observe objects for which the use of JWST time cannot be justified. ESA’s Ariel M4 mission will conduct a survey of around a thousand exoplanet atmospheres from 2029 onwards, with a wavelength range extending to 7.8\,$\mu$m. The results of the Twinkle exoplanet survey will be highly valuable for informing the Ariel target selection and thus increasing the overall science yield of the mission \cite{tinetti_ariel,tinetti_ariel2,ariel_target_list}. Additionally, Twinkle's low resolution spectroscopy from space will complement the high resolution spectroscopy being conducted from the ground \cite{brogi_2017,kesseli_w76,prinoth_w189,yan_k20}.

\subsection{Solar System Survey}

Twinkle will also conduct a survey of objects within our Solar System. The rapid pointing and tracking capabilities of the spacecraft will enable the observation of objects including asteroids, comets, the outer planets and their moons. Twinkle is capable of providing an infrared spectroscopic population study of asteroids and comets to study their surface composition \cite{edwards_twinkle_ast}, following up on hundreds of targets detected by surveys such as the Vera C. Rubin Observatory's Legacy Survey of Space and Time (LSST)  \cite{solontoi,schwamb_lsst} and the Near-Earth Object Surveyor Mission (NEOSM) \cite{mainzer_neosm}. Furthermore, Twinkle will be able to obtain high-SNR spectra of major solar system bodies within very brief exposure times \cite{edwards_twinkle_solar}. Its wavelength coverage and position above the atmosphere make it particularly well-suited for studying infrared spectral features that are obscured by telluric lines from the ground, including hydration features, organics, silicates and CO$_{\rm 2}$.

\section{Current and Future Work}

BSSL, Airbus and ABB are currently further detailing the design of the Twinkle mission. Key ongoing work includes refining the payload interfaces, the sizing of the field and cold stops, and complex thermal analysis of the spacecraft. BSSL is continuing to develop cutting-edge software to model scientific performance of the engineering design for scientists to interpret which will further enable the design to be thoroughly tested and validated. This focus on providing the tools needed by scientists to model the performance of the mission for their science cases of interest is fundamental to the design process and helps build on the industry specialisms responsible for spacecraft design.

The members of Twinkle's extrasolar survey are developing ideas on the ways in which the spacecraft can be utilised and are defining the survey's major science themes. Additionally, the Twinkle Science Team is, in co-ordination with BSSL, helping to develop the operational structure and policies of the survey. These will ensure that the surveys are undertaken with a collaborative spirit that maximises their efficiencies and provides opportunities for early career researchers to be at the forefront of the science definition and, once in orbit, data exploitation.

\clearpage

\bibliography{main} 

\begin{thebibliography}{10}

\bibitem{archer_bssl}
{Archer}, R., {Tessenyi}, M., {Tinetti}, G., {Tennyson}, J., {Faulkes}, M.~C.,
  {Savini}, G., {Windred}, P., {Brown}, D., {Edwards}, B., {Stotesbury}, I.,
  {Joshua}, M., and {Wilcock}, B., ``{A sustainable path for space science},''
  {\em Nature Astronomy}~{\bf 4},  1017--1018 (Nov. 2020).

\bibitem{boley_new_space}
{Boley}, A.~C., {Kendall}, D., {Byers}, M., {Grandmont}, F.~J., {Byers}, C.,
  {Busler}, J., {Evans}, M., {Gladman}, B., {Harrison}, T., and {Johnson}, C.,
  ``{The Role of NewSpace in Furthering Canadian Astronomy},'' in [{\em
  Canadian Long Range Plan for Astronomy and Astrophysics White
  Papers}{\nolinebreak\hspace{0.1em}]},   {\bf 2020},  4 (Oct. 2019).

\bibitem{serjeant_small_sat}
Serjeant, S., Elvis, M., and Tinetti, G., ``The future of astronomy with small
  satellites,'' {\em Nature Astronomy}~{\bf 4}(11),  1031--1038 (2020).

\bibitem{savini}
Savini, G., Tessenyi, M., Tinetti, G., Arena, C., Tennyson, J., Zingales, T.,
  Pascale, E., Sudiwala, R., Papageorgiou, A., Sarkar, S., Ade, P., Griffin,
  M.~J., Barnes, K., Hipwood, L., Knowles, P., Patel, M., Leese, M., Mason, J.,
  Crook, M., Department, T., Tosh, I., Saad, A., Eccleston, P., Shaughnessy,
  B., Brooke, T., Wells, M., Bryson, I., MacLeod, A., Taylor, W., Bezawada, N.,
  Wright, G.~S., Jason, S., Friend, J., Williams, J., Johnston, G., Prasad, S.,
  Vora, A., Saunders, C., Winter, B., Curry, P., and Smith, A., ``{Twinkle –
  A Low Earth Orbit Visible and Infrared Exoplanet Spectroscopy Observatory},''
  in [{\em Preprint (SPIE astronomical telescopes + instrumentation 2016 paper
  9904—175)}{\nolinebreak\hspace{0.1em}]},  (2016).

\bibitem{jason}
Jason, S., da~Silva~Curiel, A., Tessenyi, M., Tinetti, G., Savini, G.,
  Tennyson, J., Pascale, E., Williams, Johnson, J., Prasad, S., Vora, A.,
  Saunders, C., Friend, J., and Sweeting, M., ``Twinkle: A new idea for
  commercial astrophysics missions,'' {\em 4S Symposium, Valletta, Malta, 30
  May 2016} (2016).

\bibitem{edwards_twinkle_exo}
{Edwards}, B., {Rice}, M., {Zingales}, T., {Tessenyi}, M., {Waldmann}, I.,
  {Tinetti}, G., {Pascale}, E., {Savini}, G., and {Sarkar}, S., ``{Exoplanet
  spectroscopy and photometry with the Twinkle space telescope},'' {\em
  Experimental Astronomy}~{\bf 47},  29--63 (Apr 2019).

\bibitem{edwards_twinkle_ast}
{Edwards}, B., {Lindsay}, S., {Savini}, G., {Tinetti}, G., {Arena}, C.,
  {Bowles}, N., and {Tessenyi}, M., ``{Small bodies science with the Twinkle
  space telescope},'' {\em Journal of Astronomical Telescopes, Instruments, and
  Systems}~{\bf 5},  034004 (Jul 2019).

\bibitem{edwards_twinkle_solar}
{Edwards}, B., {Savini}, G., {Tinetti}, G., {Tessenyi}, M., {Arena}, C.,
  {Lindsay}, S., and {Bowles}, N., ``{Remote-sensing characterization of major
  Solar System bodies with the Twinkle space telescope},'' {\em Journal of
  Astronomical Telescopes, Instruments, and Systems}~{\bf 5},  014006 (Jan.
  2019).

\bibitem{edwards_terminus}
{Edwards}, B. and {Stotesbury}, I., ``{Terminus: A Versatile Simulator for
  Space-based Telescopes},'' {\em \aj}~{\bf 161},  266 (June 2021).

\bibitem{benz_cheops}
{Benz}, W., {Broeg}, C., {Fortier}, A., {Rand o}, N., {Beck}, T., {Beck}, M.,
  {Queloz}, D., {Ehrenreich}, D., {Maxted}, P., {Isaak}, K., {Billot}, N.,
  {Alibert}, Y., {Alonso}, R., {Ant{\'o}nio}, C., {Asquier}, J., {Bandy}, T.,
  {B{\'a}rczy}, T., {Barrado}, D., {Barros}, S., {Baumjohann}, W., {Bekkelien},
  A., {Bergomi}, M., {Biondi}, F., {Bonfils}, X., {Borsato}, L., {Brand eker},
  A., {Busch}, M.-D., {Cabrera}, J., {Cessa}, V., {Charnoz}, S., {Chazelas},
  B., {Collier Cameron}, A., {Corral Van Damme}, C., {Cortes}, D., {Davies},
  M., {Deleuil}, M., {Deline}, A., {Delrez}, L., {Demangeon}, O., {Demory},
  B.-O., {Erikson}, A., {Farinato}, J., {Fossati}, L., {Fridlund}, M.,
  {Futyan}, D., {Gand olfi}, D., {Garcia Munoz}, A., {Gillon}, M., {Guterman},
  P., {Gutierrez}, A., {Hasiba}, J., {Heng}, K., {Hernandez}, E., {Hoyer}, S.,
  {Kiss}, L., {Kovacs}, Z., {Kuntzer}, T., {Laskar}, J., {Lecavelier des
  Etangs}, A., {Lendl}, M., {L{\'o}pez}, A., {Lora}, I., {Lovis}, C.,
  {L{\"u}ftinger}, T., {Magrin}, D., {Malvasio}, L., {Marafatto}, L.,
  {Michaelis}, H., {de Miguel}, D., {Modrego}, D., {Munari}, M., {Nascimbeni},
  V., {Olofsson}, G., {Ottacher}, H., {Ottensamer}, R., {Pagano}, I.,
  {Palacios}, R., {Pall{\'e}}, E., {Peter}, G., {Piazza}, D., {Piotto}, G.,
  {Pizarro}, A., {Pollaco}, D., {Ragazzoni}, R., {Ratti}, F., {Rauer}, H.,
  {Ribas}, I., {Rieder}, M., {Rohlfs}, R., {Safa}, F., {Salatti}, M., {Santos},
  N., {Scandariato}, G., {S{\'e}gransan}, D., {Simon}, A., {Smith}, A.,
  {Sordet}, M., {Sousa}, S., {Steller}, M., {Szab{\'o}}, G., {Szoke}, J.,
  {Thomas}, N., {Tschentscher}, M., {Udry}, S., {Van Grootel}, V., {Viotto},
  V., {Walter}, I., {Walton}, N., {Wildi}, F., and {Wolter}, D., ``{The CHEOPS
  mission},'' {\em arXiv e-prints} ,  arXiv:2009.11633 (Sept. 2020).

\bibitem{arielrad}
{Mugnai}, L.~V., {Pascale}, E., {Edwards}, B., {Papageorgiou}, A., and
  {Sarkar}, S., ``{ArielRad: the Ariel radiometric model},'' {\em Experimental
  Astronomy}~{\bf 50},  303--328 (Oct. 2020).

\bibitem{tinetti_ariel}
{Tinetti}, G., {Drossart}, P., {Eccleston}, P., {Hartogh}, P., {Heske}, A.,
  {Leconte}, J., {Micela}, G., {Ollivier}, M., {Pilbratt}, G., {Puig}, L.,
  {Turrini}, D., {Vandenbussche}, B., {Wolkenberg}, P., {Beaulieu}, J.-P.,
  {Buchave}, L.~A., {Ferus}, M., {Griffin}, M., {Guedel}, M., {Justtanont}, K.,
  {Lagage}, P.-O., {Machado}, P., {Malaguti}, G., {Min}, M.,
  {N{\o}rgaard-Nielsen}, H.~U., {Rataj}, M., {Ray}, T., {Ribas}, I., {Swain},
  M., {Szabo}, R., {Werner}, S., {Barstow}, J., {Burleigh}, M., {Cho}, J., {du
  Foresto}, V.~C., {Coustenis}, A., {Decin}, L., {Encrenaz}, T., {Galand}, M.,
  {Gillon}, M., {Helled}, R., {Morales}, J.~C., {Mu{\~n}oz}, A.~G., {Moneti},
  A., {Pagano}, I., {Pascale}, E., {Piccioni}, G., {Pinfield}, D., {Sarkar},
  S., {Selsis}, F., {Tennyson}, J., {Triaud}, A., {Venot}, O., {Waldmann}, I.,
  {Waltham}, D., {Wright}, G., {Amiaux}, J., {Augu{\`e}res}, J.-L.,
  {Berth{\'e}}, M., {Bezawada}, N., {Bishop}, G., {Bowles}, N., {Coffey}, D.,
  {Colom{\'e}}, J., {Crook}, M., {Crouzet}, P.-E., {Da Peppo}, V., {Sanz},
  I.~E., {Focardi}, M., {Frericks}, M., {Hunt}, T., {Kohley}, R., {Middleton},
  K., {Morgante}, G., {Ottensamer}, R., {Pace}, E., {Pearson}, C., {Stamper},
  R., {Symonds}, K., {Rengel}, M., {Renotte}, E., {Ade}, P., {Affer}, L.,
  {Alard}, C., {Allard}, N., {Altieri}, F., {Andr{\'e}}, Y., {Arena}, C.,
  {Argyriou}, I., {Aylward}, A., {Baccani}, C., {Bakos}, G., {Banaszkiewicz},
  M., {Barlow}, M., {Batista}, V., {Bellucci}, G., {Benatti}, S., {Bernardi},
  P., {B{\'e}zard}, B., {Blecka}, M., {Bolmont}, E., {Bonfond}, B., {Bonito},
  R., {Bonomo}, A.~S., {Brucato}, J.~R., {Brun}, A.~S., {Bryson}, I., {Bujwan},
  W., {Casewell}, S., {Charnay}, B., {Pestellini}, C.~C., {Chen}, G.,
  {Ciaravella}, A., {Claudi}, R., {Cl{\'e}dassou}, R., {Damasso}, M.,
  {Damiano}, M., {Danielski}, C., {Deroo}, P., {Di Giorgio}, A.~M., {Dominik},
  C., {Doublier}, V., {Doyle}, S., {Doyon}, R., {Drummond}, B., {Duong}, B.,
  {Eales}, S., {Edwards}, B., {Farina}, M., {Flaccomio}, E., {Fletcher}, L.,
  {Forget}, F., {Fossey}, S., {Fr{\"a}nz}, M., {Fujii}, Y.,
  {Garc{\'\i}a-Piquer}, {\'A}., {Gear}, W., {Geoffray}, H., {G{\'e}rard},
  J.~C., {Gesa}, L., {Gomez}, H., {Graczyk}, R., {Griffith}, C., {Grodent}, D.,
  {Guarcello}, M.~G., {Gustin}, J., {Hamano}, K., {Hargrave}, P., {Hello}, Y.,
  {Heng}, K., {Herrero}, E., {Hornstrup}, A., {Hubert}, B., {Ida}, S., {Ikoma},
  M., {Iro}, N., {Irwin}, P., {Jarchow}, C., {Jaubert}, J., {Jones}, H.,
  {Julien}, Q., {Kameda}, S., {Kerschbaum}, F., {Kervella}, P., {Koskinen}, T.,
  {Krijger}, M., {Krupp}, N., {Lafarga}, M., {Landini}, F., {Lellouch}, E.,
  {Leto}, G., {Luntzer}, A., {Rank-L{\"u}ftinger}, T., {Maggio}, A.,
  {Maldonado}, J., {Maillard}, J.-P., {Mall}, U., {Marquette}, J.-B., {Mathis},
  S., {Maxted}, P., {Matsuo}, T., {Medvedev}, A., {Miguel}, Y., {Minier}, V.,
  {Morello}, G., {Mura}, A., {Narita}, N., {Nascimbeni}, V., {Nguyen Tong}, N.,
  {Noce}, V., {Oliva}, F., {Palle}, E., {Palmer}, P., {Pancrazzi}, M.,
  {Papageorgiou}, A., {Parmentier}, V., {Perger}, M., {Petralia}, A.,
  {Pezzuto}, S., {Pierrehumbert}, R., {Pillitteri}, I., {Piotto}, G., {Pisano},
  G., {Prisinzano}, L., {Radioti}, A., {R{\'e}ess}, J.-M., {Rezac}, L.,
  {Rocchetto}, M., {Rosich}, A., {Sanna}, N., {Santerne}, A., {Savini}, G.,
  {Scandariato}, G., {Sicardy}, B., {Sierra}, C., {Sindoni}, G., {Skup}, K.,
  {Snellen}, I., {Sobiecki}, M., {Soret}, L., {Sozzetti}, A., {Stiepen}, A.,
  {Strugarek}, A., {Taylor}, J., {Taylor}, W., {Terenzi}, L., {Tessenyi}, M.,
  {Tsiaras}, A., {Tucker}, C., {Valencia}, D., {Vasisht}, G., {Vazan}, A.,
  {Vilardell}, F., {Vinatier}, S., {Viti}, S., {Waters}, R., {Wawer}, P.,
  {Wawrzaszek}, A., {Whitworth}, A., {Yung}, Y.~L., {Yurchenko}, S.~N.,
  {Osorio}, M. R.~Z., {Zellem}, R., {Zingales}, T., and {Zwart}, F., ``{A
  chemical survey of exoplanets with ARIEL},'' {\em Experimental
  Astronomy}~{\bf 46},  135--209 (Nov. 2018).

\bibitem{tinetti_ariel2}
{Tinetti}, G., {Eccleston}, P., {Haswell}, C., {Lagage}, P.-O., {Leconte}, J.,
  {L{\"u}ftinger}, T., {Micela}, G., {Min}, M., {Pilbratt}, G., {Puig}, L.,
  {Swain}, M., {Testi}, L., {Turrini}, D., {Vandenbussche}, B., {Rosa Zapatero
  Osorio}, M., {Aret}, A., {Beaulieu}, J.-P., {Buchhave}, L., {Ferus}, M.,
  {Griffin}, M., {Guedel}, M., {Hartogh}, P., {Machado}, P., {Malaguti}, G.,
  {Pall{\'e}}, E., {Rataj}, M., {Ray}, T., {Ribas}, I., {Szab{\'o}}, R., {Tan},
  J., {Werner}, S., {Ratti}, F., {Scharmberg}, C., {Salvignol}, J.-C.,
  {Boudin}, N., {Halain}, J.-P., {Haag}, M., {Crouzet}, P.-E., {Kohley}, R.,
  {Symonds}, K., {Renk}, F., {Caldwell}, A., {Abreu}, M., {Alonso}, G.,
  {Amiaux}, J., {Berth{\'e}}, M., {Bishop}, G., {Bowles}, N., {Carmona}, M.,
  {Coffey}, D., {Colom{\'e}}, J., {Crook}, M., {D{\'e}sjonqueres}, L.,
  {D{\'\i}az}, J.~J., {Drummond}, R., {Focardi}, M., {G{\'o}mez}, J.~M.,
  {Holmes}, W., {Krijger}, M., {Kovacs}, Z., {Hunt}, T., {Machado}, R.,
  {Morgante}, G., {Ollivier}, M., {Ottensamer}, R., {Pace}, E., {Pagano}, T.,
  {Pascale}, E., {Pearson}, C., {M{\o}ller Pedersen}, S., {Pniel}, M., {Roose},
  S., {Savini}, G., {Stamper}, R., {Szirovicza}, P., {Szoke}, J., {Tosh}, I.,
  {Vilardell}, F., {Barstow}, J., {Borsato}, L., {Casewell}, S., {Changeat},
  Q., {Charnay}, B., {Civi{\v{s}}}, S., {Coud{\'e} du Foresto}, V.,
  {Coustenis}, A., {Cowan}, N., {Danielski}, C., {Demangeon}, O., {Drossart},
  P., {Edwards}, B.~N., {Gilli}, G., {Encrenaz}, T., {Kiss}, C., {Kokori}, A.,
  {Ikoma}, M., {Morales}, J.~C., {Mendon{\c{c}}a}, J., {Moneti}, A., {Mugnai},
  L., {Garc{\'\i}a Mu{\~n}oz}, A., {Helled}, R., {Kama}, M., {Miguel}, Y.,
  {Nikolaou}, N., {Pagano}, I., {Panic}, O., {Rengel}, M., {Rickman}, H.,
  {Rocchetto}, M., {Sarkar}, S., {Selsis}, F., {Tennyson}, J., {Tsiaras}, A.,
  {Venot}, O., {Vida}, K., {Waldmann}, I.~P., {Yurchenko}, S., {Szab{\'o}}, G.,
  {Zellem}, R., {Al-Refaie}, A., {Perez Alvarez}, J., {Anisman}, L.,
  {Arhancet}, A., {Ateca}, J., {Baeyens}, R., {Barnes}, J.~R., {Bell}, T.,
  {Benatti}, S., {Biazzo}, K., {B{\l}{\k{e}}cka}, M., {Bonomo}, A.~S., {Bosch},
  J., {Bossini}, D., {Bourgalais}, J., {Brienza}, D., {Brucalassi}, A.,
  {Bruno}, G., {Caines}, H., {Calcutt}, S., {Campante}, T., {Canestrari}, R.,
  {Cann}, N., {Casali}, G., {Casas}, A., {Cassone}, G., {Cara}, C., {Carmona},
  M., {Carone}, L., {Carrasco}, N., {Changeat}, Q., {Chioetto}, P.,
  {Cortecchia}, F., {Czupalla}, M., {Chubb}, K.~L., {Ciaravella}, A., {Claret},
  A., {Claudi}, R., {Codella}, C., {Garcia Comas}, M., {Cracchiolo}, G.,
  {Cubillos}, P., {Da Peppo}, V., {Decin}, L., {Dejabrun}, C., {Delgado-Mena},
  E., {Di Giorgio}, A., {Diolaiti}, E., {Dorn}, C., {Doublier}, V.,
  {Doumayrou}, E., {Dransfield}, G., {Dumaye}, L., {Dunford}, E., {Jimenez
  Escobar}, A., {Van Eylen}, V., {Farina}, M., {Fedele}, D., {Fern{\'a}ndez},
  A., {Fleury}, B., {Fonte}, S., {Fontignie}, J., {Fossati}, L., {Funke}, B.,
  {Galy}, C., {Garai}, Z., {Garc{\'\i}a}, A., {Garc{\'\i}a-Rigo}, A., {Garufi},
  A., {Germano Sacco}, G., {Giacobbe}, P., {G{\'o}mez}, A., {Gonzalez}, A.,
  {Gonzalez-Galindo}, F., {Grassi}, D., {Griffith}, C., {Guarcello}, M.~G.,
  {Goujon}, A., {Gressier}, A., {Grzegorczyk}, A., {Guillot}, T., {Guilluy},
  G., {Hargrave}, P., {Hellin}, M.-L., {Herrero}, E., {Hills}, M., {Horeau},
  B., {Ito}, Y., {Jessen}, N.~C., {Kabath}, P., {K{\'a}lm{\'a}n}, S.,
  {Kawashima}, Y., {Kimura}, T., {Kn{\'\i}{\v{z}}ek}, A., {Kreidberg}, L.,
  {Kruid}, R., {Kruijssen}, D. J.~M., {Kubel{\'\i}k}, P., {Lara}, L.,
  {Lebonnois}, S., {Lee}, D., {Lefevre}, M., {Lichtenberg}, T., {Locci}, D.,
  {Lombini}, M., {Sanchez Lopez}, A., {Lorenzani}, A., {MacDonald}, R.,
  {Magrini}, L., {Maldonado}, J., {Marcq}, E., {Migliorini}, A.,
  {Modirrousta-Galian}, D., {Molaverdikhani}, K., {Molinari}, S.,
  {Molli{\`e}re}, P., {Moreau}, V., {Morello}, G., {Morinaud}, G., {Morvan},
  M., {Moses}, J.~I., {Mouzali}, S., {Nakhjiri}, N., {Naponiello}, L.,
  {Narita}, N., {Nascimbeni}, V., {Nikolaou}, A., {Noce}, V., {Oliva}, F.,
  {Palladino}, P., {Papageorgiou}, A., {Parmentier}, V., {Peres}, G.,
  {P{\'e}rez}, J., {Perez-Hoyos}, S., {Perger}, M., {Cecchi Pestellini}, C.,
  {Petralia}, A., {Philippon}, A., {Piccialli}, A., {Pignatari}, M., {Piotto},
  G., {Podio}, L., {Polenta}, G., {Preti}, G., {Pribulla}, T., {Lopez Puertas},
  M., {Rainer}, M., {Reess}, J.-M., {Rimmer}, P., {Robert}, S., {Rosich}, A.,
  {Rossi}, L., {Rust}, D., {Saleh}, A., {Sanna}, N., {Schisano}, E.,
  {Schreiber}, L., {Schwartz}, V., {Scippa}, A., {Seli}, B., {Shibata}, S.,
  {Simpson}, C., {Shorttle}, O., {Skaf}, N., {Skup}, K., {Sobiecki}, M.,
  {Sousa}, S., {Sozzetti}, A., {{\v{S}}poner}, J., {Steiger}, L., {Tanga}, P.,
  {Tackley}, P., {Taylor}, J., {Tecza}, M., {Terenzi}, L., {Tremblin}, P.,
  {Tozzi}, A., {Triaud}, A., {Trompet}, L., {Tsai}, S.-M., {Tsantaki}, M.,
  {Valencia}, D., {Carine Vandaele}, A., {Van der Swaelmen}, M., {Vardan}, A.,
  {Vasisht}, G., {Vazan}, A., {Del Vecchio}, C., {Waltham}, D., {Wawer}, P.,
  {Widemann}, T., {Wolkenberg}, P., {Hou Yip}, G., {Yung}, Y., {Zilinskas}, M.,
  {Zingales}, T., and {Zuppella}, P., ``{Ariel: Enabling planetary science
  across light-years},'' {\em arXiv e-prints} ,  arXiv:2104.04824 (Apr. 2021).

\bibitem{Exosim}
{Sarkar}, S., {Pascale}, E., {Papageorgiou}, A., {Johnson}, L.~J., and
  {Waldmann}, I., ``{ExoSim: the Exoplanet Observation Simulator},'' {\em arXiv
  e-prints} ,  arXiv:2002.03739 (Feb. 2020).

\bibitem{kuntzer_salsa}
{Kuntzer}, T., {Fortier}, A., and {Benz}, W., ``{SALSA: a tool to estimate the
  stray light contamination for low-Earth orbit observatories},'' in [{\em
  \procspie}{\nolinebreak\hspace{0.1em}]},  {\em Society of Photo-Optical
  Instrumentation Engineers (SPIE) Conference Series} {\bf 9149},  91490W (Aug.
  2014).

\bibitem{swain_finesse}
{Deroo}, P., {Swain}, M.~R., and {Green}, R.~O., ``{Spectroscopy of exoplanet
  atmospheres with the FINESSE Explorer mission},'' in [{\em
  \procspie}{\nolinebreak\hspace{0.1em}]},  {\em Society of Photo-Optical
  Instrumentation Engineers (SPIE) Conference Series} {\bf 8442},  844241
  (Sept. 2012).

\bibitem{JosephK2-266}
{Rodriguez}, J.~E., {Becker}, J.~C., {Eastman}, J.~D., {Hadden}, S.,
  {Vanderburg}, A., {Khain}, T., {Quinn}, S.~N., {Mayo}, A., {Dressing}, C.~D.,
  {Schlieder}, J.~E., {Ciardi}, D.~R., {Latham}, D.~W., {Rappaport}, S.,
  {Adams}, F.~C., {Berlind}, P., {Bieryla}, A., {Calkins}, M.~L., {Esquerdo},
  G.~A., {Kristiansen}, M.~H., {Omohundro}, M., {Schwengeler}, H.~M.,
  {Stassun}, K.~G., and {Terentev}, I., ``{A Compact Multi-planet System with a
  Significantly Misaligned Ultra Short Period Planet},'' {\em \aj}~{\bf 156},
  245 (Nov. 2018).

\bibitem{rodriguez_k2_266}
{Rodriguez}, J.~E., {Becker}, J.~C., {Eastman}, J.~D., {Hadden}, S.,
  {Vanderburg}, A., {Khain}, T., {Quinn}, S.~N., {Mayo}, A., {Dressing}, C.~D.,
  {Schlieder}, J.~E., {Ciardi}, D.~R., {Latham}, D.~W., {Rappaport}, S.,
  {Adams}, F.~C., {Berlind}, P., {Bieryla}, A., {Calkins}, M.~L., {Esquerdo},
  G.~A., {Kristiansen}, M.~H., {Omohundro}, M., {Schwengeler}, H.~M.,
  {Stassun}, K.~G., and {Terentev}, I., ``{A Compact Multi-planet System with a
  Significantly Misaligned Ultra Short Period Planet},'' {\em \aj}~{\bf 156},
  245 (Nov. 2018).

\bibitem{ercolano_twinkle_pah}
{Ercolano}, B., {Rab}, C., {Molaverdikhani}, K., {Edwards}, B., {Preibisch},
  T., {Testi}, L., {Kamp}, I., and {Thi}, W.-F., ``{Observations of PAHs in the
  atmospheres of discs and exoplanets},'' {\em \mnras}~{\bf 512},  430--438
  (May 2022).

\bibitem{tsiaras_plc}
{Tsiaras}, A., {Waldmann}, I., {Rocchetto}, M., {Varley}, R., {Morello}, G.,
  {Damiano}, M., and {Tinetti}, G., ``{pylightcurve: Exoplanet lightcurve
  model},''  ascl:1612.018 (Dec. 2016).

\bibitem{seok_pah}
Seok, J.~Y. and Li, A., ``Polycyclic aromatic hydrocarbons in protoplanetary
  disks around herbig ae/be and t tauri stars,'' {\em The Astrophysical
  Journal}~{\bf 835},  291 (feb 2017).

\bibitem{oberg_2011}
{{\"O}berg}, K.~I., {Murray-Clay}, R., and {Bergin}, E.~A., ``{The Effects of
  Snowlines on C/O in Planetary Atmospheres},'' {\em \apjl}~{\bf 743},  L16
  (Dec. 2011).

\bibitem{mordasini_2016}
{Mordasini}, C., {van Boekel}, R., {Molli{\`e}re}, P., {Henning}, T., and
  {Benneke}, B., ``{The Imprint of Exoplanet Formation History on Observable
  Present-day Spectra of Hot Jupiters},'' {\em \apj}~{\bf 832},  41 (Nov.
  2016).

\bibitem{apai_tlse}
{Apai}, D., {Rackham}, B.~V., {Giampapa}, M.~S., {Angerhausen}, D., {Teske},
  J., {Barstow}, J., {Carone}, L., {Cegla}, H., {Domagal-Goldman}, S.~D.,
  {Espinoza}, N., {Giles}, H., {Gully-Santiago}, M., {Haywood}, R., {Hu}, R.,
  {Jordan}, A., {Kreidberg}, L., {Line}, M., {Llama}, J., {L{\'o}pez-Morales},
  M., {Marley}, M.~S., and {de Wit}, J., ``{Understanding Stellar Contamination
  in Exoplanet Transmission Spectra as an Essential Step in Small Planet
  Characterization},'' {\em arXiv e-prints} ,  arXiv:1803.08708 (Mar. 2018).

\bibitem{apai_bd}
{Apai}, D., {Radigan}, J., {Buenzli}, E., {Burrows}, A., {Reid}, I.~N., and
  {Jayawardhana}, R., ``{HST Spectral Mapping of L/T Transition Brown Dwarfs
  Reveals Cloud Thickness Variations},'' {\em \apj}~{\bf 768},  121 (May 2013).

\bibitem{biller_bd}
{Biller}, B.~A., {Vos}, J., {Buenzli}, E., {Allers}, K., {Bonnefoy}, M.,
  {Charnay}, B., {B{\'e}zard}, B., {Allard}, F., {Homeier}, D., {Bonavita}, M.,
  {Brandner}, W., {Crossfield}, I., {Dupuy}, T., {Henning}, T., {Kopytova}, T.,
  {Liu}, M.~C., {Manjavacas}, E., and {Schlieder}, J., ``{Simultaneous
  Multiwavelength Variability Characterization of the Free-floating
  Planetary-mass Object PSO J318.5-22},'' {\em \aj}~{\bf 155},  95 (Feb. 2018).

\bibitem{bowler_2020}
{Bowler}, B.~P., {Zhou}, Y., {Morley}, C.~V., {Kataria}, T., {Bryan}, M.~L.,
  {Benneke}, B., and {Batygin}, K., ``{Strong Near-infrared Spectral
  Variability of the Young Cloudy L Dwarf Companion VHS J1256-1257 b},'' {\em
  \apjl}~{\bf 893},  L30 (Apr. 2020).

\bibitem{jiang_w43}
{Jiang}, I.-G., {Lai}, C.-Y., {Savushkin}, A., {Mkrtichian}, D., {Antonyuk},
  K., {Griv}, E., {Hsieh}, H.-F., and {Yeh}, L.-C., ``{The Possible Orbital
  Decay and Transit Timing Variations of the Planet WASP-43b},'' {\em \aj}~{\bf
  151},  17 (Jan. 2016).

\bibitem{turner_w12}
{Turner}, J.~D., {Ridden-Harper}, A., and {Jayawardhana}, R., ``{Decaying Orbit
  of the Hot Jupiter WASP-12b: Confirmation with TESS Observations},'' {\em
  \aj}~{\bf 161},  72 (Feb. 2021).

\bibitem{wong_w12}
{Wong}, I., {Shporer}, A., {Vissapragada}, S., {Greklek-McKeon}, M., {Knutson},
  H.~A., {Winn}, J.~N., and {Benneke}, B., ``{TESS Revisits WASP-12: Updated
  Orbital Decay Rate and Constraints on Atmospheric Variability},'' {\em
  \aj}~{\bf 163},  175 (Apr. 2022).

\bibitem{tsiaras_pop}
{Tsiaras}, A., {Waldmann}, I.~P., {Zingales}, T., {Rocchetto}, M., {Morello},
  G., {Damiano}, M., {Karpouzas}, K., {Tinetti}, G., {McKemmish}, L.~K.,
  {Tennyson}, J., and {Yurchenko}, S.~N., ``{A Population Study of Gaseous
  Exoplanets},'' {\em \aj}~{\bf 155},  156 (Apr. 2018).

\bibitem{wallack_spitzer_pop}
{Wallack}, N.~L., {Knutson}, H.~A., {Morley}, C.~V., {Moses}, J.~I., {Thomas},
  N.~H., {Thorngren}, D.~P., {Deming}, D., {D{\'e}sert}, J.-M., {Fortney},
  J.~J., and {Kammer}, J.~A., ``{Investigating Trends in Atmospheric
  Compositions of Cool Gas Giant Planets Using Spitzer Secondary Eclipses},''
  {\em \aj}~{\bf 158},  217 (Dec. 2019).

\bibitem{baxter_spitzer_pop}
{Baxter}, C., {D{\'e}sert}, J.-M., {Tsai}, S.-M., {Todorov}, K.~O., {Bean},
  J.~L., {Deming}, D., {Parmentier}, V., {Fortney}, J.~J., {Line}, M.,
  {Thorngren}, D., {Pierrehumbert}, R.~T., {Burrows}, A., and {Showman}, A.~P.,
  ``{Evidence for disequilibrium chemistry from vertical mixing in hot Jupiter
  atmospheres. A comprehensive survey of transiting close-in gas giant
  exoplanets with warm-Spitzer/IRAC},'' {\em \aap}~{\bf 648},  A127 (Apr.
  2021).

\bibitem{changeat_emission_pop}
{Changeat}, Q., {Edwards}, B., {Al-Refaie}, A.~F., {Tsiaras}, A., {Skinner},
  J.~W., {Cho}, J.~Y.~K., {Yip}, K.~H., {Anisman}, L., {Ikoma}, M., {Bieger},
  M.~F., {Venot}, O., {Shibata}, S., {Waldmann}, I.~P., and {Tinetti}, G.,
  ``{Five Key Exoplanet Questions Answered via the Analysis of 25 Hot-Jupiter
  Atmospheres in Eclipse},'' {\em \apjs}~{\bf 260},  3 (May 2022).

\bibitem{keating_hierarchical}
{Keating}, D. and {Cowan}, N.~B., ``{Atmospheric characterization of hot
  Jupiters using hierarchical models of Spitzer observations},'' {\em
  \mnras}~{\bf 509},  289--299 (Jan. 2022).

\bibitem{bieryla_k7}
{Bieryla}, A., {Collins}, K., {Beatty}, T.~G., {Eastman}, J., {Siverd}, R.~J.,
  {Pepper}, J., {Gaudi}, B.~S., {Stassun}, K.~G., {Ca{\~n}as}, C., {Latham},
  D.~W., {Buchhave}, L.~A., {Sanchis-Ojeda}, R., {Winn}, J.~N., {Jensen}, E.
  L.~N., {Kielkopf}, J.~F., {McLeod}, K.~K., {Gregorio}, J., {Col{\'o}n},
  K.~D., {Street}, R., {Ross}, R., {Penny}, M., {Mellon}, S.~N., {Oberst},
  T.~E., {Fulton}, B.~J., {Wang}, J., {Berlind}, P., {Calkins}, M.~L.,
  {Esquerdo}, G.~A., {DePoy}, D.~L., {Gould}, A., {Marshall}, J., {Pogge}, R.,
  {Trueblood}, M., and {Trueblood}, P., ``{KELT-7b: A Hot Jupiter Transiting a
  Bright V = 8.54 Rapidly Rotating F-star},'' {\em \aj}~{\bf 150},  12 (July
  2015).

\bibitem{pluriel_k7}
{Pluriel}, W., {Whiteford}, N., {Edwards}, B., {Changeat}, Q., {Yip}, K.~H.,
  {Baeyens}, R., {Al-Refaie}, A., {Fabienne Bieger}, M., {Blain}, D.,
  {Gressier}, A., {Guilluy}, G., {Yassin Jaziri}, A., {Kiefer}, F.,
  {Modirrousta-Galian}, D., {Morvan}, M., {Mugnai}, L.~V., {Poveda}, M.,
  {Skaf}, N., {Zingales}, T., {Wright}, S., {Charnay}, B., {Drossart}, P.,
  {Leconte}, J., {Tsiaras}, A., {Venot}, O., {Waldmann}, I., and {Beaulieu},
  J.-P., ``{ARES. III. Unveiling the Two Faces of KELT-7 b with HST WFC3},''
  {\em \aj}~{\bf 160},  112 (Sept. 2020).

\bibitem{taurex3}
{Al-Refaie}, A.~F., {Changeat}, Q., {Waldmann}, I.~P., and {Tinetti}, G.,
  ``{TauREx 3: A Fast, Dynamic, and Extendable Framework for Retrievals},''
  {\em \apj}~{\bf 917},  37 (Aug. 2021).

\bibitem{rothman_hitremp_2010}
Rothman, L., Gordon, I., Barber, R., Dothe, H., Gamache, R., Goldman, A.,
  Perevalov, V., Tashkun, S., and Tennyson, J., ``Hitemp, the high-temperature
  molecular spectroscopic database,'' {\em Journal of Quantitative Spectroscopy
  and Radiative Transfer}~{\bf 111}(15),  2139--2150 (2010).

\bibitem{polyansky_h2o}
Polyansky, O.~L., Kyuberis, A.~A., Zobov, N.~F., Tennyson, J., Yurchenko,
  S.~N., and Lodi, L., ``Exomol molecular line lists xxx: a complete
  high-accuracy line list for water,'' {\em Monthly Notices of the Royal
  Astronomical Society}~{\bf 480}(2),  2597--2608 (2018).

\bibitem{ExoMol_CH4_new}
Yurchenko, S.~N., Amundsen, D.~S., Tennyson, J., and Waldmann, I.~P., ``{A
  hybrid line list for CH$_4$ and hot methane continuum },'' {\em A\&A}~{\bf
  605},  A95 (2017).

\bibitem{li_co_2015}
Li, G., Gordon, I.~E., Rothman, L.~S., Tan, Y., Hu, S.-M., Kassi, S.,
  Campargue, A., and Medvedev, E.~S., ``Rovibrational line lists for nine
  isotopologues of the co molecule in the x 1$\sigma$+ ground electronic
  state,'' {\em The Astrophysical Journal Supplement Series}~{\bf 216}(1),  15
  (2015).

\bibitem{ariel_target_list}
{Edwards}, B. and {Tinetti}, G., ``{The Ariel Target List: The Impact of TESS
  and the Potential for Characterizing Multiple Planets within a System},''
  {\em \aj}~{\bf 164},  15 (July 2022).

\bibitem{brogi_2017}
Brogi, M., Line, M., Bean, J., D{\'{e}}sert, J.-M., and Schwarz, H., ``A
  framework to combine low- and high-resolution spectroscopy for the
  atmospheres of transiting exoplanets,'' {\em The Astrophysical Journal}~{\bf
  839},  L2 (apr 2017).

\bibitem{kesseli_w76}
{Kesseli}, A.~Y., {Snellen}, I.~A.~G., {Casasayas-Barris}, N., {Molli{\`e}re},
  P., and {S{\'a}nchez-L{\'o}pez}, A., ``{An Atomic Spectral Survey of
  WASP-76b: Resolving Chemical Gradients and Asymmetries},'' {\em \aj}~{\bf
  163},  107 (Mar. 2022).

\bibitem{prinoth_w189}
{Prinoth}, B., {Hoeijmakers}, H.~J., {Kitzmann}, D., {Sandvik}, E., {Seidel},
  J.~V., {Lendl}, M., {Borsato}, N.~W., {Thorsbro}, B., {Anderson}, D.~R.,
  {Barrado}, D., {Kravchenko}, K., {Allart}, R., {Bourrier}, V., {Cegla},
  H.~M., {Ehrenreich}, D., {Fisher}, C., {Lovis}, C., {Guzm{\'a}n-Mesa}, A.,
  {Grimm}, S., {Hooton}, M., {Morris}, B.~M., {Oreshenko}, M., {Pino}, L., and
  {Heng}, K., ``{Titanium oxide and chemical inhomogeneity in the atmosphere of
  the exoplanet WASP-189 b},'' {\em Nature Astronomy}~{\bf 6},  449--457 (Jan.
  2022).

\bibitem{yan_k20}
{Yan}, F., {Reiners}, A., {Pall{\'e}}, E., {Shulyak}, D., {Stangret}, M.,
  {Molaverdikhani}, K., {Nortmann}, L., {Molli{\`e}re}, P., {Henning}, T.,
  {Casasayas-Barris}, N., {Cont}, D., {Chen}, G., {Czesla}, S.,
  {S{\'a}nchez-L{\'o}pez}, A., {L{\'o}pez-Puertas}, M., {Ribas}, I.,
  {Quirrenbach}, A., {Caballero}, J.~A., {Amado}, P.~J.,
  {Galad{\'\i}-Enr{\'\i}quez}, D., {Khalafinejad}, S., {Lara}, L.~M., {Montes},
  D., {Morello}, G., {Nagel}, E., {Sedaghati}, E., {Zapatero Osorio}, M.~R.,
  and {Zechmeister}, M., ``{Detection of iron emission lines and a temperature
  inversion on the dayside of the ultra-hot Jupiter KELT-20b},'' {\em
  \aap}~{\bf 659},  A7 (Mar. 2022).

\bibitem{solontoi}
{Solontoi}, M., {Ivezi{\'c}}, {\v Z}., and {Jones}, L., ``{Comet Science with
  LSST},'' in [{\em American Astronomical Society Meeting Abstracts
  \#217}{\nolinebreak\hspace{0.1em}]},  {\em Bulletin of the American
  Astronomical Society} {\bf 43},  252.11 (Jan. 2011).

\bibitem{schwamb_lsst}
{Schwamb}, M.~E., {Jones}, R.~L., {Chesley}, S.~R., {Fitzsimmons}, A.,
  {Fraser}, W.~C., {Holman}, M.~J., {Hsieh}, H., {Ragozzine}, D., {Thomas},
  C.~A., {Trilling}, D.~E., {Brown}, M.~E., {Bannister}, M.~T., {Bodewits}, D.,
  {de Val-Borro}, M., {Gerdes}, D., {Granvik}, M., {Kelley}, M. S.~P.,
  {Knight}, M.~M., {Seaman}, R.~L., {Ye}, Q.-Z., and {Young}, L.~A., ``{Large
  Synoptic Survey Telescope Solar System Science Roadmap},'' {\em arXiv
  e-prints} ,  arXiv:1802.01783 (Feb. 2018).

\bibitem{mainzer_neosm}
{Mainzer}, A., {Abell}, P., {Bauer}, J., {Bottke}, W., {Grav}, T., {Kelley},
  M., {Kramer}, E., {Masci}, F., {Masiero}, J., {Reddy}, V., {Reinhart}, L.,
  {Sonnett}, S., {Wright}, E., and {Wong}, A., ``{Near-Earth Object Surveyor
  Mission: Data Products and Survey Plan},'' in [{\em AAS/Division for
  Planetary Sciences Meeting Abstracts}{\nolinebreak\hspace{0.1em}]},  {\em
  AAS/Division for Planetary Sciences Meeting Abstracts} {\bf 53},  306.16
  (Oct. 2021).

\end{thebibliography}
\bibliographystyle{spiebib} 

\end{document}